# Sharing Experience around Component Compositions – Application to the Eclipse Ecosystem

Grégory Bourguin[1,2], Arnaud Lewandowski[1,2], and Myriam Lewkowicz[3]
[1] Univ Lille Nord de France, France
[2] ULCO, LISIC, Calais, France
[3] Troyes University of Technology, ICD/Tech-CICO, Troyes, France

**ABSTRACT**

We nowadays live in a world of tailorable systems in which end-users are able to transform their working environment while achieving their tasks, day to day and over the time. Tailorability is most of the time achieved through dynamic component integration thanks to a huge number of components available over the Internet. In this context, the main problem for users is not anymore the integration of new components, but how to find the most interesting set of components that will fulfill their needs. Facing this issue, our assumption is that it would be helpful for users to take benefit of the experience of other users and our work aims at enhancing current software ecosystems to support this sharing of experience. We have applied this approach in the context of software development while considering Eclipse as one of the most advanced and used software ecosystem. We then offer ShareXP, an Eclipse feature that allows members of a group to share their expertise, this expertise being embodied in the "compositions" each of them has built. ShareXP was already presented in (Bourguin et al., 2012). The current paper is an extension where we deeper show that ShareXP is only a first step in our global approach trying to enhance not only the Eclipse ecosystem, but software ecosystems in general.

**INTRODUCTION**

Tailorability is today much more than a research concept and many of the currently widely used software are in fact tailorable. This statement can be verified in many different application domains: for instance image editing with Photoshop and its plug-ins, Business Process Management (BPM) with solutions offering diverse connectors for service integration, computer games with World of Warcraft and the success of its add-ons which enable users to enhance their playing environment, and, of course, software development with the Eclipse ecosystem and its "everything is a plug-in" approach (Gamma and Beck, 2004). Even if solutions exist and are used in everyday life, there are remaining issues, and a large part of the research work in end-user development tries to propose better tailorable systems (Zhu et al., 2011). Not surprisingly, most of this research is directed towards new means for integration, since this way of tailoring has long been identified as offering a good equilibrium between the level of tailorability that can be achieved, and the level of expertise or effort that is required to achieve it (Mørch, 1997). A lot of work still remains for understanding how to facilitate this component integration.

However, many systems already propose well-defined integration means, and there is a real huge collection of components that can be easily downloaded and integrated. In this context, the matter for users is not anymore how to integrate new components, but how to find the most interesting sets of components that will fulfill their needs. Facing this issue of finding and selecting good components for performing particular tasks, we suggest that it would be helpful for users to take benefit of the experience of other users.

In fact, the individual skills of users and their ability to share and generate knowledge within their communities and social networks play a crucial role for organizations which want to continuously learn. Networks of personal relationships which are created and reinforced through interpersonal conversation are critical in supporting knowledge sharing (Erickson and Kellogg, 2002). As an example, this has been showed in a field study on the use of Eclipse done by Draxler et al. (2011a) where they found that integration work in the Eclipse ecosystem is a social activity where actors rely on their local social network. The diffusion or sharing is often rooted in personal contacts, project teams, work groups or even a whole organization. But what is the knowledge that users of the Eclipse ecosystem could share for their organization to learn? Finding a component does not give any knowledge by itself. In fact, the context of its use makes the difference. And context arises from embodied knowledge, which is information that is uniquely and integrally embodied in the person's personality, creativity, intelligence, perceptions, experiences and relationships (Fitzpatrick, 2002). Embodied knowledge is the essence of expertise. It is difficult to identify a priori what expertise could be shared, but it comes naturally in the course of a conversation, in response to hearing or seeing a connected theme and choosing to contribute (Fitzpatrick, 2002). In tailorable systems, components assemblages are elements where knowledge is embodied. A particular assemblage or composition represents a particular point of view on the working environment that best suits a specific task. We will show in part 2 how, when faced with new needs, or as their expertise grows, users can modify their compositions: they add new components, modify their graphical arrangement, etc. Compositions thus crystallize theirs users experience while reflecting some part of their expertise, as they are the result of users' embodied knowledge while realizing their tasks. Better supporting the sharing such particular compositions is our proposition for expertise sharing in tailorable environments.

In order to test our ideas, we have particularly developed this approach in the context of Eclipse. Eclipse is one of the most used, most tailorable, and most studied environments where a particular components composition is called a perspective (Springgay, 2001). As a result, we offer ShareXP, a sociotechnical system in the sense that we aim at designing usability while supporting sociability (Preece, 2000), and that was presented in (Bourguin et al., 2012). By using ShareXP, users are able to browse particular composition(s) created by others, to preview a composition according to a chosen perspective, to chat about them, and eventually to integrate a full composition or some of its (sub-)component(s) in their own environment. By offering these functionalities, we follow three of the guidelines defined by Wulf et al. (2008) to improve component-based tailorability: (1) We offer an "exploration environment" allowing the simulation of the interface, (2) the shared repository (containing the components) is integrated into the tailoring environment, (3) the shared repository is activated directly with only those components that are relevant for a certain tailoring context being displayed (the perspectives).

This paper is an extension of (Bourguin et al., 2012) where we will show that ShareXP is only a first step towards our main objective: trying to enhance experience sharing not only in the Eclipse ecosystem, but in software ecosystems in general. The paper is organized as follow: after a presentation of existing software ecosystems and the need to support the sharing of components compositions, we will focus on Eclipse, considering it as one of the most used and advanced software ecosystem. We will then position ourselves according to related work on sharing components compositions in the context of the Eclipse Ecosystem. We will then present ShareXP, the Eclipse feature that we offer. We will finally discuss the limits and future work.

## EXISTING SOFTWARE ECOSYSTEMS

Considering large and tailorable environments, Messerschmitt and Szyperski (2003) introduced the term of software ecosystem to reflect the network of actors – software vendors, developers, and users – interacting in a social, organizational, or technical environment. Software ecosystem has also been defined as "a collection of software projects, which are developed and which co-evolve in the same environment" (Lungu, 2009). The relationships between the many actors of software ecosystems has also been emphasized by Draxler et al. (2011a): software ecosystems "consist of an open, extensible software platform that attracts different manufacturers and hobbyists, creating small-scale components, which can be individually assembled by end-users".

We summarize the interactions that can be generally observed between different parts and actors of common Software Ecosystems in figure 1: *developers* deliver *Components* or (predefined) *Components Compositions* that are stored or referenced in *Components Repositories*. *Users* are able to browse *Components Repositories* to discover, install and compose the (set of) *components* that will fulfill their needs.

*Figure 1. Interactions between actors of a software ecosystem*

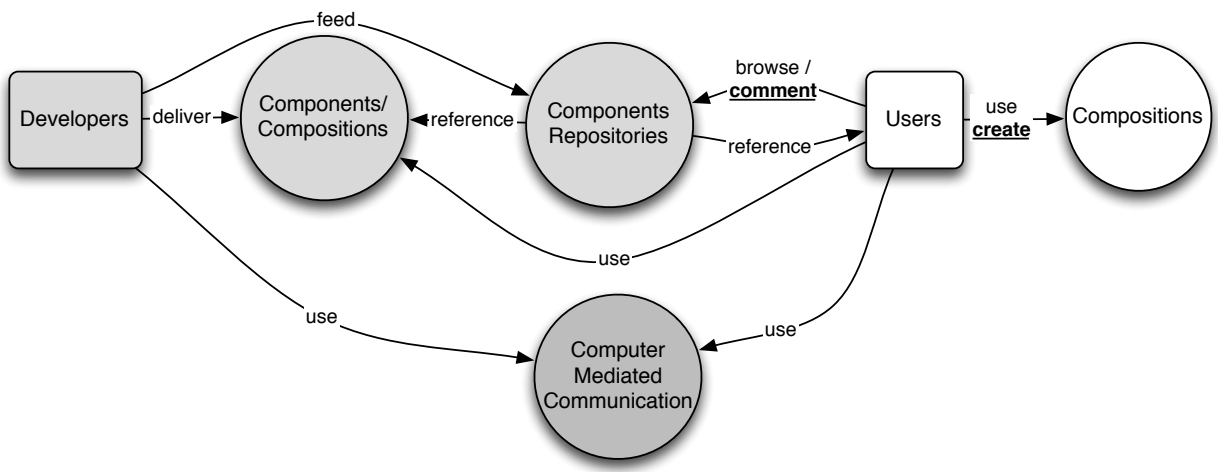

As an example, we can consider the domain of software development and Eclipse that constitutes one of the most famous software ecosystems. Eclipse end-users are mainly software developers organized in teams involved in the realization of complex and multi-facets projects. Eclipse is a platform that offers radical component based tailorability by following the "everything is a plug-in" approach (Gamma and Beck, 2004). According to this principle, the platform itself consists in sets of plug-ins called features[1] that take in charge the basic mechanisms for loading plug-ins, starting them, etc. On this core platform, all the functionalities that may be proposed by Eclipse stand as components (features). Eclipse then proposes different predefined components compositions or packages that are dedicated to different purposes; for example, we can find Eclipse IDE for Java Developers, Eclipse Modeling Tools, Eclipse for Testers, etc.

Besides this rich diversity, the openness of the platform has led to the emergence of a huge amount of third-party components (i.e. features not proposed nor developed by Eclipse, but by other users/organizations) that can be downloaded by Eclipse end-users in order to customize their working environment and dynamically adapt it to their needs. These features are available through components repositories like the Eclipse MarketPlace[2].

*Figure 2. Components (features) and Component Compositions (perspectives) in Eclipse*

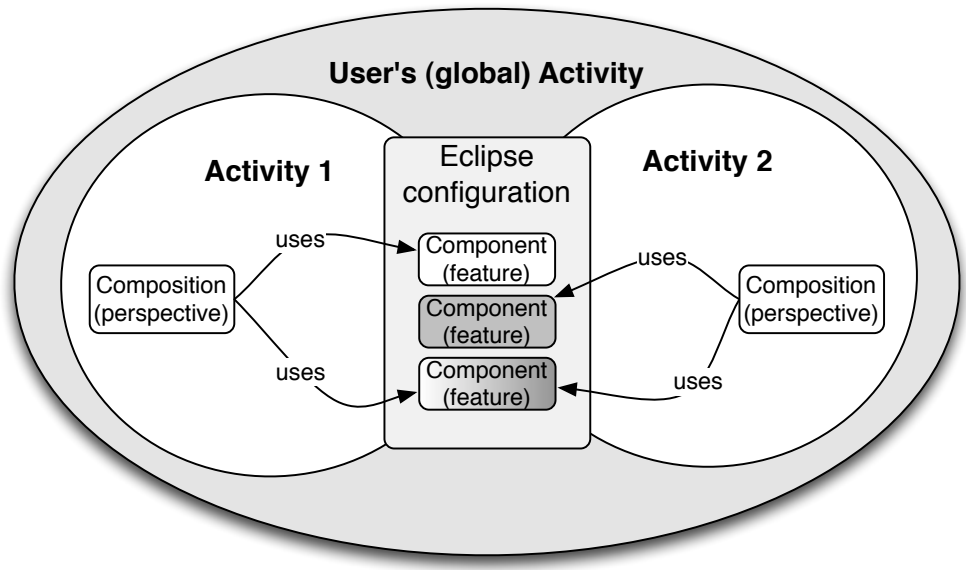

As a result, Eclipse users usually install many different features dedicated to their different tasks. However, if many features are then available in their Eclipse configuration, users do not use all of them at all time. As illustrated on figure 2, only a subset of features is involved in a particular task. In Eclipse, such a subset, a particular composition of features, corresponds to the *perspective* core concept. A perspective controls the presence of elements provided by features and that contribute to the user's interface (windows, action buttons, etc.) (Springgay, 2001). The purpose of a perspective is to present a selection of the functionalities that may be useful in a specific activity. Users can have several perspectives, one for each task they have to perform. Some perspectives are included in Eclipse packages, others are provided with specific complex features. For example, the Java perspective defines a set of tools (package explorer, code editor, etc) that are useful for Java development. One perspective is active at a time, and users can easily switch from one perspective to another as the focus of their work changes. But perspectives are not frozen. A user can adapt and/or create perspectives by modifying their elements (components to be shown, placements in the interface, available actions in the toolbars, etc.) and thus develop its own components compositions.

---

[1] A feature is a logical unit defining a set of plug-ins (and eventually other features) working together, and on which it depends.
[2] Eclipse Marketplace, http://marketplace.eclipse.org/

The Eclipse MarketPlace offers more than 1300 different downloadable features. In this context, the main matter for end-users is how to discover the most interesting set of components that will fulfill their needs. In order to cope with this problem, the components are categorized regarding the tasks they are supposed to support. The Eclipse MarketPlace is structured around 47 categories in which each component is presented with a small description and screenshot(s). It is also possible to store and show eventual users' comments or reviews about each component. Another interesting section is the "Favorited by" part that shows a list of names that reference other users that have marked a solution as one of their preferred one.

The same mechanisms can be found in many similar software ecosystems dedicated to different other domains. We already talked about World of Warcraft and its 10 millions of subscribers who mainly play their roles while using sets of third parties add-ons developed to enhance and personalize their gaming experience. These can for example be discovered through the Curse[3] repository that proposes to find, download, comment and favorite 5705 WoW add-ons and 2908 WoW add-ons packs (predefined compositions of add-ons). Browsing Curse reveals that this repository also centralizes Rift[4] addons, Warhammer Online[5] addons, Minecraft[6] mods and more in the same way. More generally, one can notice that the 1 billion smartphone users (Ramanathan, 2012) discover, download, comment and favorite their component applications through similar repositories like the iTunes App Store and the Android Market.

If these repositories and their functionalities are indisputably helpful for users, we will show deeper in the next section that they have revealed to be not sufficient since most of the sharing of experience between users about their "useful" set of components remains realized through separate computer-mediated communication (CMC) tools, and in an ad hoc way.

## THE NEED FOR SHARING CONTEXTUALIZED EXPERTISE

As an example, we can consider what happened a few time ago among a small team of researchers who were developing a prototype in our own laboratory. The prototype was developed using Eclipse and no collaborative tool was used apart the email and Subversion[7] for sharing its source code.

One of the developers (Peter) needed to build some user interfaces in Java using the SWT toolkit. Many Eclipse features are available on the Internet for this purpose but he did not know which one to choose for performing his task. Peter knew he may find advices on forums or sites such as Stackoverflow[8] by asking questions such as: "which Eclipse tool can help me to create Graphical User Interfaces?[9]" but remembered that a colleague (John) involved in the same project had already performed this kind of task (GUI design). Peter went to John's desk and asked him which tool(s) he should use. John opened the adequate perspective in Eclipse and briefly showed the development environment he is using for GUI design. Peter saw a tool that could be very useful for his task and which could correspond to his style of working. He pointed the desired component to John's screen and asked him for its name. John searched the feature name through the list of his installed features, and sent a small mail to Peter. Back to his desk, Peter installed the feature, and faced many difficulties to use it properly. It even did not look like in John's perspective. Peter went then back to John's desk to discuss the problem, and they finally discovered that John made a mistake: he transmitted the name of another GUI design feature. The problem came from the fact that John had previously tested several features dedicated to GUI design before finding his preferred one, and even if they were not currently used, they were still present in its (technical) working environment (they did not the trick in this project, but may reveal helpful for another one...).

---

[3] Curse, http://www.curse.com/
[4] Rift, http://eu.riftgame.com/
[5] Warhammer Online, http://warhammeronline.com/
[6] Minecraft, https://minecraft.net/
[7] Apache Subversion, http://subversion.apache.org/

This example illustrates several needs. Firstly, Peter preferred asking John for advices about features to use rather than searching for the good one on the marketplace. He may have posted his question on Stackoverflow to have an idea about the features commonly used by other fellow programmers, but the finding of the "good" feature on the Internet is a hard task: in fact, before John made his choice, he himself had to browse through several features and preferred testing them rather than simply trust forums, reviews and notations displayed on the Internet, even if these latter gave him interesting (first) indications about their quality. Secondly, the visual aspect of the feature, and its integration in someone else's environment played an important role in helping Peter to imagine how this feature could fit his own working environment. Finally, the main computer mediated communication tool they have used was the email system, which is definitely not dedicated to Eclipse components sharing.

These needs around experience and expertise sharing are aligned with those presented in (Draxler et al., 2011a) and (Stevens and Draxler, 2010), based on their field study of teams of software developers using the Eclipse ecosystem. One of their issue was "How and based on what information do people modify their personal software installation?". Strongly inspired by the work of Mackay (1990), they underline that this evolution mainly results from end-users "collegial" collaboration. For example, the authors report a usual situation they call a kind of "initiation rite" in which experienced senior Eclipse developers would tell a new team member what features to install by passing the features names or even the whole set of artefacts to the newcomer. As a result, these developers worked for a certain time with the same configuration, and their environment often later drifted apart as each one modified its own assembling according to its own experience while realizing its tasks. The authors report another example where Eclipse users sat together at one machine to discuss a problem. They show that the discovery of the colleague's environment also often "accidentally" leads to an exchange of potential interesting features for performing their task.

---

[8] http://stackoverflow.com is a programming Question&Answer website
[9] http://stackoverflow.com/questions/6533243/create-gui-using-eclipse-java

More generally, the study presented in (Draxler et al., 2011b) shows mechanisms about how end-users manage what they call "software portfolios". The authors compare practices of developers managing their set of Eclipse plug-ins or features, with practices of players managing their set of World of Warcraft add-ons. They demonstrate that in both worlds, users rely on the recommendations of other people to compose their environment, and they trust advices coming from their well-known colleagues or friends much more than "anonymous" recommendations found on the Internet or magazines. Finally, a particular composition is often shared among users who are involved in the same kind of task. The studies conclude by pointing "a clear lack of tools for supporting the various practices of collaboration with regard to appropriating/tailoring".

*Figure 3. Proposition for enhancing users' experience sharing inside software ecosystems*

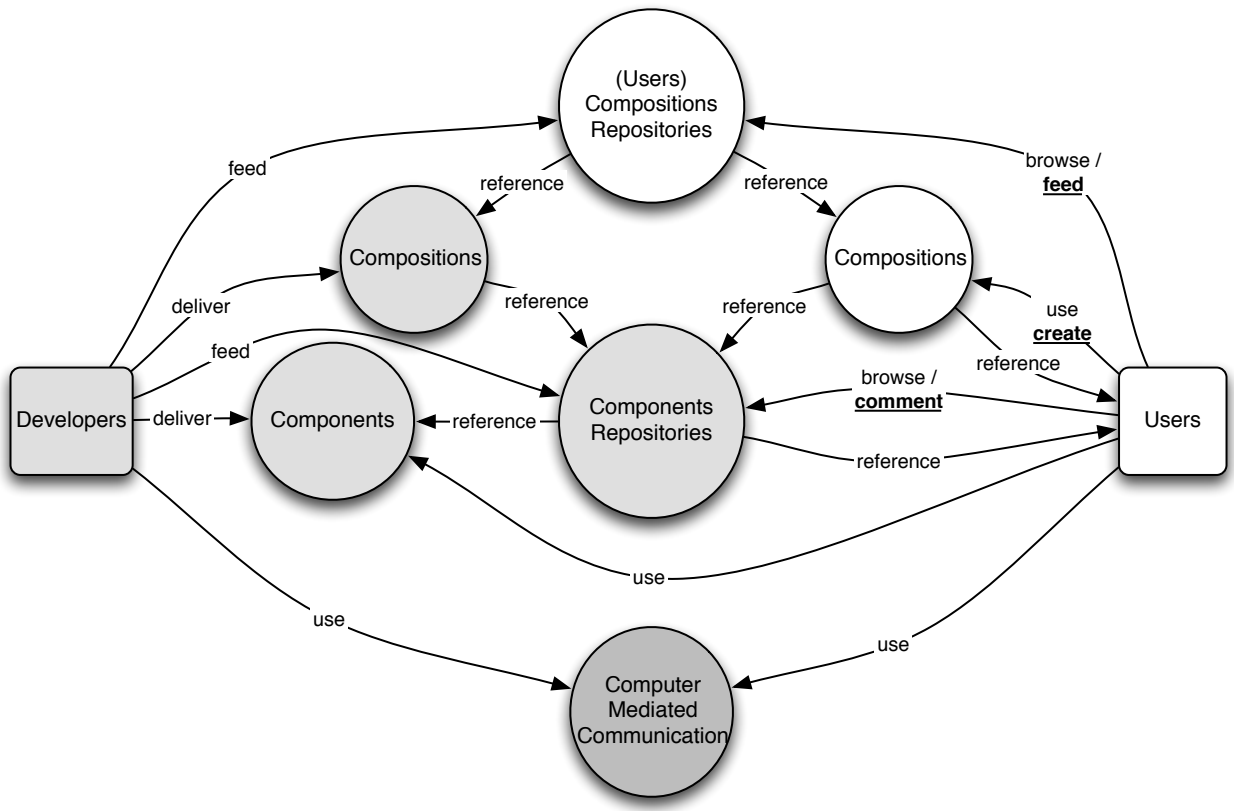

As we described in the previous section, a part of this lack resides in the fact that current software ecosystems offer poor dedicated support for users in order to help them share their experience. Existing components repositories set "isolated" components as first class objects. As such, the component constitutes the main entry point of these repositories. Users' experience can only be found by selecting a particular component, and browsing its accompanying users' comments. We can hardly find any comments on a specific composition (a perspective) since perspectives are not directly referenced by these repositories. Components are categorized regarding the tasks they are supposed to support. One can notice that research studies have for long shown that (computer) tools are often not used as they were supposed to be. This also raises the question of the tasks grain that is considered. Users may share through CMC tools like forums or email systems but no tool reifies components compositions as first class objects or entities that can directly be managed and shared. As a result, the experience that is crystallized in end-users' components compositions is still mainly shared in an ad hoc fashion.

To overcome this problem, Figure 3 shows how our approach aims at enhancing software ecosystems by offering new dedicated tools for sharing end-users experience through **compositions** repositories. The main idea is to complement existing software ecosystems by proposing a reverse view in which the main entry point is not the component, but the end-user's task(s) s/he is involved in. In this approach, users will be directly able to browse particular components compositions that have been developed by others in their concrete tasks, thus taking benefit from their experience that has been developed while performing their day-to-day activities.

Following this approach in the context of the Eclipse ecosystem, we have created ShareXP, a new tool in which users are able (1) to browse the knowledge embodied in particular features composition(s) created by others while realizing their tasks, (2) to discuss about that, and eventually (3) to integrate a full composition or some of its (sub-)component(s) in their own environment.

## RELATED WORK

The need for expertise sharing in the Eclipse platform has been studied by researchers and software companies which propose external solutions to share or to be able to contextualize the features.

### Existing Solution for Sharing Features

Peerclipse (Draxler et al., 2009) is a research prototype. It allows sharing and retrieving Eclipse configurations in a local community. The set of features available in the configuration of each member is shared over the community through JXTA, and other members are then able to browse and install them in their own environment.

However, Eclipse users usually manage around 40 features in their Eclipse configuration. Moreover, a usual scenario may imply a novice user who needs to browse an expert configuration for discovering helpful tools. As an expertized user is usually involved in different parts of a complex project, or in many different projects, there is a high probability that the Eclipse configuration of this expert contains much more than 40 features. For instance, Stevens and Draxler (2010) found a maximum of 196 features in a single Eclipse installation. Indeed, browsing such consequent set of features for finding the one(s) that could help in the realization of a particular task may also reveal a complex and painful activity.

A platform like Eclipse is usually used for several projects. Users develop their environment over a long period and the resulting configuration comes from their implication in diverse tasks. Thus, as illustrated before on figure 2 (section 2), each installed feature may participate to some of the user's activities, but certainly does not participate to all of them. In other words, each of the 40 features found in an Eclipse configuration does not participate in each activity that is or has been realized by its user.

It may sometimes be instructive to discover the whole environment of a user who has developed a certain expertise in a particular task. However, the main goal of a novice user is usually to perform his/her specific task, and it is in this context that s/he would like to benefit from an expertise. The whole listing of tools/features installed by the expert while realizing tasks with no direct link with the novice's one is not interesting at that time. It can even reveal to be counterproductive by submerging the novice user under a huge quantity of information without a direct link with the sought one. To sum up, "simply" browsing the set of features installed in an Eclipse configuration does not permit to provide the useful information which is the context of use of each feature.

From our point of view, the remaining questions are: for which task(s) my colleague used this feature? Which other features did s/he use when solving this task? And eventually, how did s/he assemble these features together? The answers remain in what we called a components composition that describes the (sub)set of features involved in the context of a specific user task.

### Existing Solution for Contextualizing Features

Yoxos[10] is a commercial product which aims at managing several Eclipse workspaces by introducing the Yoxos profile concept. A Yoxos profile is usually dedicated to a specific task and contains a specific list of features that should be used for executing a specific task. The Yoxos launcher lets users select the profile they want to start and the tool configures an Eclipse workspace according to it, with only the needed features. Users may have access to different profiles to be able to start working with different environments according to specific tasks. When a user starts a profile referencing a feature that is missing in his own environment, the feature is dynamically downloaded and installed through the Yoxos certified shared repository of components that provides the same kind of service than the Eclipse Marketplace.

Yoxos users have a direct access to the Yoxos repository where they can find components (like in the Marketplace) and can integrate them in their own profiles, thus making them evolve. Yoxos profiles can be shared within the whole Yoxos community, or between the members of a team. As a result, Yoxos users can browse the profiles of other users, discover the list of features involved in each one, decide to install specific components, or decide to install the whole profile.

As profiles are usually created by expert users and dedicated to specific tasks, this approach helps to discover a set of features from other users in their specific context of use. However, a set of features is mainly described by text; there is no exploration environment that presents the integration of a set of components at the user interface level. It is then less obvious for a user to really understand the context of use of this set of components. As we pointed out in our example scenario and as it was already shown by (Draxler, 2011b), a user usually asks for expertise sharing when looking at the working environment of the expert.

### SHAREXP

As we described in section 2, a perspective represents the setup of a user and provides links to the list of features involved in a particular task, links between these features for realizing this task, and even the placement of the tools at the Graphical User Interface (GUI) level. As Eclipse perspectives can be created and adapted by end-users while performing their task, they are key elements which embody the expertise of the end-users that they have acquired during their past experiences. Drawing from this assumption, we offer ShareXP, a feature prototype that supports expertise sharing through perspective sharing (Bourguin et al., 2012).

---

[10] Yoxos, http://eclipsesource.com/en/yoxos/

ShareXP has been realized as a standard Eclipse feature that can be installed/started by users to enhance their development environment. Following the main trend in Eclipse collaborative development tools like Jazz (Hupfer et al., 2004), ShareXP has been built over the Eclipse Communication Framework (ECF). As a result, ShareXP may well integrate or complement Eclipse workspaces that cover other collaborative dimensions thanks to tools like those offered by Jazz.

ShareXP relies on the eXtensible Messaging and Presence Protocol (XMPP) so that XMPP users do not have to install additional server: they can thus for example use their existing gmail account for sharing with their existing contacts (including their team fellows). The ShareXP contacts view (see figure 4) shows whether other users are online or not. Users can discuss together using their usual chat tool/plug-in, or decide to use ShareXP's included one. As ShareXP uses the existing user's XMPP account, using the included chat plug-in does not involve the creation of a new discussion channel. This approach prevents from introducing a split of the conversation thread between different tools.

*Figure 4. View showing user's contacts and giving access to their expertise*

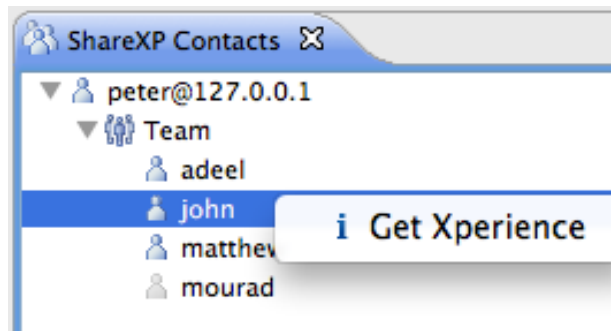

Back to our example scenario presented in section 3, Peter and John have now integrated ShareXP in their working environment. Peter knows that John has already tested some features for GUI development. Thanks to ShareXP, he can select his colleague in the contact list to discover his shared perspectives (NB: using ShareXP's preferences panel, users can decide not to share their own configurations). To share the perspectives of a user, ShareXP serializes their internal description, along with a screen capture of each one. An object representing these perspectives is then sent to the network so that every contact asking for the perspectives of this fellow can get the object describing his perspectives. Figure 5 shows how Peter is able to browse the list of John's perspectives (part (a)). When Peter selects GUI Development, he can see a preview of this perspective (part (b)). As the mouse flies over the different parts of this preview, the wizard highlights the pointed components, their corresponding names and the feature they are packed with(part (c)). Some of the features used by John in this particular task may not provide a specific GUI. Peter can also discover them in part (d) that presents the whole set of features involved in the selected perspective.

*Figure 5. Browsing John's perspectives (a) – with preview (b,c) and related features (d) – helps Peter to find the appropriate one for his task*

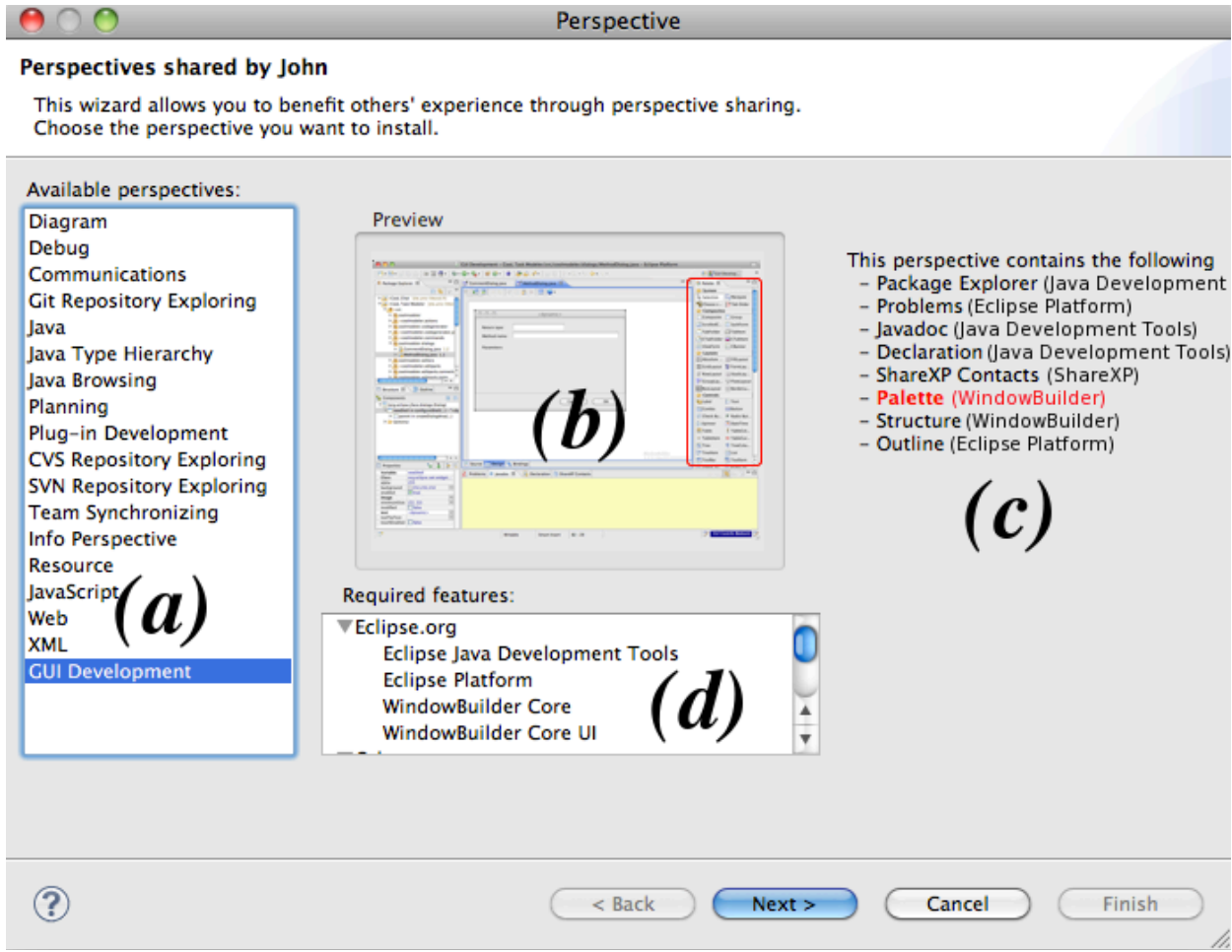

Once Peter has found the component(s) that best suits his needs, his next step (figure 6) is to select what to install in his own environment. He can select some specific features, or decide to install the whole perspective, including the graphical arrangement of each view, actions toolbars, etc. created by John. ShareXP checks if the features are already available in the local environment, along with their version, and everything is then automatically installed. This installation is performed through the Eclipse public API dedicated to plug-in installation (Equinox/p2) for the features, and eventually by copying John's perspective in the local environment.

*Figure 6. Select which features to install along with John's perspective*

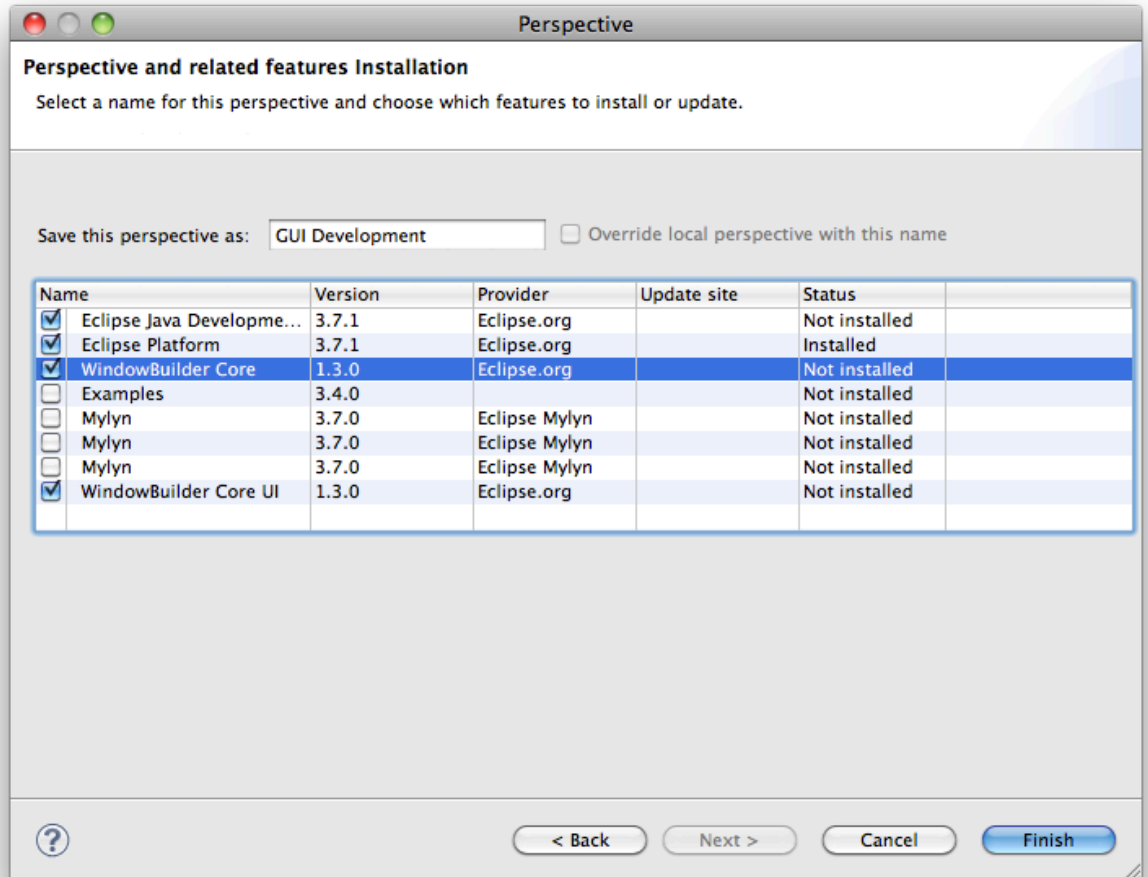

This approach presents three benefits. First, ShareXP allows users to browse and discuss others (whole) configurations filtered by perspectives, so including the context of work. Only the features used in the selected perspective are shown: Peter can only see the features used by John in the specific context of the GUI Development task. This prevents him to have to find his way through a set of many components that are out of his scope. Secondly, the approach encourages the discovering of unexpected features which are directly related to the context of work; As Peter is browsing John's GUI Development perspective, he may discover other types of features (useful for this task) next to the ones he was looking for. Finally, Peter benefits from John's expertise in GUI Development even at the GUI level: the visual arrangement of the useful features is written in the shared perspective; windows are placed at the right places, and toolbars are configured ideally according to John's experience.

## LIMITS AND FUTURE WORK

The solutions described in this paper, including those presented in the related work, were made possible thanks to Eclipse's good concepts and properties regarding tailorability. As we already said, it is important to note that Eclipse is one of the most advanced existing software ecosystem. If the issues presented in this paper are common to most software ecosystems, they do not all currently offer the concepts and mechanisms that we needed to implement our solution. In ShareXP, our proposition mainly relies on the integrated dynamic feature installation mechanism, and the important perspective core concept. In particular, perspectives really helped us in setting end-users components compositions as first class objects that can be managed and shared. Similar properties do not all exist in other software ecosystem. As an example, if researchers have already identified World of Warcraft as a truly interesting software ecosystem, installing new add-ons into it has to be realized out of the game, requiring a quit and restart, and no end-user concept offers a clear representation of the add-ons compositions involved by a player for performing a particular task. As a result, a lot of work still needs to be done in order to generalize the solution offered by ShareXP in the context of other software ecosystems.

Moreover, we would like to underline that, in Eclipse, even if components compositions are reified and defined at the end-users' abstraction level as perspectives, they programmatically remain internal objects, and no public API is proposed to access their entire definition. We then had to go deep into the code of the platform in order to find the components and mechanisms needed for reifying and share them. In other words, Eclipse perspectives have not been totally designed for being programmatically manipulated or shared. Considering that Eclipse perspectives have revealed to be powerful means that embodies end-users expertise, it would be a great benefit if the openness of the platform was improved by considering them as first class objects which could be fully programmatically manipulated and shared like plug-ins and features are.

In this first version of ShareXP, we decided to rely on XMPP that proposes synchronous communication in a connected mode. This choice presents limitations since users can only benefit from the expertise shared by other users that are connected when the search is performed. Like for the Yoxos profiles that we presented in section 3, it would be possible to create local copies of the perspectives of distant users as soon as a member connects to the ShareXP network. It would thus be possible to browse others' configurations even after their disconnection. This kind of mechanism was firstly not in the direct scope of ShareXP since our approach was to complement other already existing synchronous collaborative tools like Jazz (Hupfer et al., 2004): we aimed at fostering discussions around personal configurations, like those starting when someone is looking "over the shoulder" of his/her colleague. We however do not forget that the human activities we try to support are distributed in space and time, and future work in ShareXP, or in the general context of software ecosystems, will thus need to try to understand and support end-users' experience sharing while deeper considering these two dimensions.

Finally, it is important to note that ShareXP offers a solution that is limited to the Eclipse ecosystem while human activities are not. As an example, a GUI designer may use Eclipse features to program an interface whose visual design is realized thanks to Photoshop tools. The components compositions created by an experimented user do usually not only involve components issued from a single software ecosystem. Human activities concurrently involve and combine many different software ecosystems into a larger global one. This is why we strongly believe that our future work will also lead us to develop our approach described in Figure 1 in the larger context of this global multifaceted software ecosystem.

## CONCLUSION

We nowadays live in a world of tailorable software ecosystems in which end-users continually evolve their own working environment while downloading and integrating third party components. Considering the wide range of possibilities and the huge set of tools and components that are available over the Internet, we showed in this paper that one of the main issues of software ecosystems is to help end-users to find the best set of components that may help them realize their tasks. Facing it, we suggested that it would be helpful for users to take benefit of the experience of other users, experience that is crystallized in the components compositions they create while performing their different activities.

We analyzed the architecture, mechanisms and properties of current widely used software ecosystems, and underlined their lacks in supporting the important sharing of end-users components compositions. This led us to propose an approach that aims at enhancing current software ecosystems by introducing components compositions repositories. This approach has been applied in the context of the Eclipse ecosystem while considering it as one of the most used, advanced, and studied software ecosystem. Analyzing related work and Eclipse's core concept and properties, we proposed ShareXP, an Eclipse feature that supports end-users features compositions sharing through the Eclipse perspective core concept.

We underlined the benefits of ShareXP: it does not only permit to find a component, but enables its discovery through the end-user's task it is involved in. ShareXP emphasizes the context of use of a specific component while allowing previewing it in a working environment. We also pointed out the limits of this solution since a lot of work still needs to be done both in Eclipse, and even more in software ecosystems in general, to better share end-users' experience by reifying components compositions as first class objects that should be directly managed, shared and browsed by end-users in their day to day activities.